\documentclass[amsmath,amssymb,aps,prb,hyperlink,reprint,showpacs,superscriptaddress]{revtex4-1}

\usepackage{graphicx}
\usepackage{soul}
\usepackage[colorlinks=true,citecolor=blue,linkcolor=blue,urlcolor=blue]{hyperref}
\usepackage[usenames]{color}
\usepackage{amsfonts}
\usepackage{dcolumn}
\usepackage{multirow}
\usepackage{times}
\usepackage{array}
\newcommand{\PreserveBackslash}[1]{\let\temp=\\#1\let\\=\temp}
\newcolumntype{C}[1]{>{\PreserveBackslash\centering}p{#1}}
\newcolumntype{R}[1]{>{\PreserveBackslash\raggedleft}p{#1}}
\newcolumntype{L}[1]{>{\PreserveBackslash\raggedright}p{#1}}
\newcolumntype{d}[1]{D{.}{.}{#1}}
\usepackage{makecell}
\usepackage[caption=false]{subfig}
\usepackage{natbib}
\setcitestyle{numbers,square}
\makeatletter
\def\NAT@def@citea{\def\@citea{\NAT@separator}}
\makeatother

\begin{document}

\title{Electron-phonon interaction in Ca$_2$N monolayer: intrinsic mobility of electrene}

\author{Xiongzhi Zeng}
\author{Songtao Zhao}
\author{Zhenyu Li}
\email[Corresponding author ]{zyli@ustc.edu.cn}
\author{Jinlong Yang}

\affiliation{Hefei National Laboratory for Physical Sciences at the Microscale, University of Science and Technology of China, Hefei, Anhui 230026, China}

\date{\vspace{0.1cm}\today}

\begin{abstract}
Electron-phonon (e-ph) interaction in Ca$_2$N monolayer, the first electrene material with two-dimensional (2D) electron gas floating in free space, is expected to be very weak and such a character can be used to design weak-scattering transport channels. Therefore, it is highly desirable to quantitatively evaluate the carrier mobility of electrene. In this study, e-ph interaction in Ca$_2$N monolayer is investigated using a precise Wannier interpolation-based first-principles technique. The calculated e-ph coupling matrix elements of Ca$_2$N monolayer are indeed small compared to other 2D materials such as graphene, which leads to an intrinsic mobility of 189 cm$^2$V$^{-1}$s$^{-1}$, much higher than those of conventional metals. Other factors affecting mobility are discussed in a comparison with graphene.  It is predicted that, based on a momentum mismatch mechanism, mobility of Ca$_2$N monolayer can be increased further to above 3000 cm$^2$V$^{-1}$s$^{-1}$ via hole doping. Our results confirm that Ca$_2$N electrene is a promising electronic material.
\end{abstract}


\maketitle
\section{INTRODUCTION}
Recently, an interesting system containing two-dimensional electron gas in free space (2DEG-FS) has been theoretically designed \cite{2014-JACS-136-13313-Zhao} and then experimentally synthesized \cite{2016-JACS-138-16089-Druffel}. Such a system is based on an electride material, dicalcium nitride (Ca$_2$N) \cite{2013-Natrue-494-336-Lee}, where electrons act as anions. When Ca$_2$N is exfoliated into monolayers, 2DEG-FS will float on both surfaces of the atomic layer, forming a so-called electrene system \cite{2017-J.Mater.Chem.C-5-11196-druffel}. Conducting electrons in Ca$_2$N monolayer mainly come from the 2DEG-FS, of which the spatial distribution is expected to minimize the electron-phonon (e-ph) interaction. On the basis of such a reasoning, Ca$_2$N monolayer is proposed to be a good material to construct ideal transport channels for the next generation of electronics \cite{2014-JACS-136-13313-Zhao,2016-Wiley-Interdiscip-Rev-Comput-Mol-Sci-6-430-Zhao}. Although conceptually attractive, the idea to use 2DEG-FS states for transport has not been quantitatively validated yet, and corresponding transport properties, such as the phonon-limited intrinsic electron mobility, remain unknown.

In many cases, e-ph interaction is studied using simple models, such as the deformation potential theory \cite{1950-Phys.Rev-80-72-Bardeen} focusing on the longitudinal-acoustic phonon mode \cite{2008-PRB-77-235305-Leu,2009-JACS-131-17728-Long,2010-NanoLett-10-869-Murphy,2011-ACSNano-5-2593-Long,2011-Adv.Mater-23-1145-Shuai,2015-NanoLett-15-2006-Morgan}. However, to accurately calculate the electron mobility of the metallic Ca$_2$N monolayer, e-ph interaction should be systematically investigated here. In principle, all e-ph coupling matrix elements can be directly calculated based on the density functional perturbation theory (DFPT) \cite{2001-Rev.Mod.Phys-73-515-Baroni}. The problem is, to obtain well converged transport properties, a very dense sampling of the Brillouin zone (BZ) is required, which usually makes the calculations intractable. To solve this problem, a Wannier function based Fourier interpolation method \cite{2007-PRB-76-165108-Giustino} can be used to calculate the e-ph interaction strength and carrier mobility \cite{2010-PRB-81-121412-Borysenko,2014-JCP-141-034704-Xi}.

In this study, we perform well converged first-principles calculations of the e-ph interaction and mobility of Ca$_2$N monolayer using density functional theory (DFT) and DFPT. Compared to graphene, the e-ph interaction matrix elements of  Ca$_2$N monolayer are much smaller. Its intrinsic phonon-limited electron mobility is 189 cm$^2$V$^{-1}$s$^{-1}$, higher than those of widely used metals such as Cu (90 cm$^2$V$^{-1}$s$^{-1}$ according to our calculations). The mobility can be further increased to about 3000 cm$^2$V$^{-1}$s$^{-1}$ via scattering suppression at a lower carrier concentration as a requirement of energy and momentum conservation. On the basis of the BCS theory, a superconducting transition at the temperature of 4.7 K induced by e-ph interactions is predicted for Ca$_2$N monolayer. Our results confirm that Ca$_2$N monolayer is a promising material for electronic applications.

\section{METHODS}

DFT and DFPT calculations were performed with the QUANTUM-ESPRESSO package \cite{2009-JPCM-21-395502-Giannozzi}. The Perdew-Burke-Ernzerhof (PBE) \cite{1996-PRL-77-3865-Perdew} exchange correlation functional at the generalized gradient approximation (GGA) level was used together with ultrasoft pseudopotentials. Energy cutoff was 150 Ry for wavefunctions and 600 Ry for charge densities. Periodic boundary condition was used with a 25 \AA\ interlayer distance to avoid artificial interactions between neighboring layers. Ground-state calculations were carried out on a regular $36 \times 36 \times 1$ mesh of $\mathbf{k}$ points with a 0.02 Ry cold smearing \cite{1999-PRL-82-3296-Marzari} applied to electron occupations. Electron group velocities were obtained with the BoltzWann package \cite{2014-Comput.Phys.Commun-185-422-Pizzi}. In phonon calculations, a $10^{-22}$ Ry convergence threshold was used for DFPT self-consistent iterations.

Transport properties can be studied by solving the Boltzman transport equation \cite{mahan2013many} and the mobility within the relaxation time approximation is expressed as
\begin{equation}\label{eqn1}
  \mu=e\frac{\sum_n\int\tau(n,\mathbf{k})v^2(n,\mathbf{k})\frac{\partial f_{n\mathbf{k}}^0}{\partial \varepsilon_{n\mathbf{k}}}d\mathbf{k}}{\sum_n\int f_{n\mathbf{k}}^0d\mathbf{k}}
\end{equation}
where $\varepsilon_{n\mathbf{k}}$ is the electron energy, $f_{n\mathbf{k}}^0$ is the electron distribution function at equilibrium, $v(n,\mathbf{k})=|1/\hbar \nabla_\mathbf{k} \varepsilon_{n\mathbf{k}}|$ is the norm of group velocity, and $\tau(n,\mathbf{k})$ is the band- and $\mathbf{k}$-dependent relaxation time of electron, reciprocal of the scattering rate $\Gamma_{n\mathbf{k}}^{e\textrm{-}ph}$. The scattering rate can be expressed by the imaginary part of  e-ph self-energy $\Gamma_{n\mathbf{k}}^{e\textrm{-}ph}=2/\hbar $ Im$\Sigma_{n\mathbf{k}}^{e\textrm{-}ph} $, and it is calculated using the following equation \cite{2016-Comput.Phys.Commun-209-116-Ponce}
\begin{multline}\label{eqn2}
  \Gamma_{n\mathbf{k}}^{e\textrm{-}ph}=\frac{2\pi}{\hbar}\sum_{m\nu\mathbf{q}}|g_{nm\nu} (\mathbf{k,q})|^2\\
  \times[(N_{\nu\mathbf{q}}^0+f_{m\mathbf{k+q}}^0)\delta(\varepsilon_{n\mathbf{k}}- \varepsilon_{m\mathbf{k+q}}+\hbar\omega_{\nu\mathbf{q}})\\
  +(N_{\nu\mathbf{q}}^0+1-f_{m\mathbf{k+q}}^0)\delta(\varepsilon_{n\mathbf{k}}- \varepsilon_{m\mathbf{k+q}}-\hbar\omega_{\nu\mathbf{q}})]
\end{multline}
where $\hbar\omega_{\nu\mathbf{q}}$ is the phonon energy, $N_{\nu\mathbf{q}}^0$ is the Bose-Einstein distribution, and $g_{nm\nu} (\mathbf{k,q})$ is the e-ph coupling matrix element corresponding to electron scattering from band $n$ at wavevector $\mathbf{k}$ to band $m$ at $\mathbf{k+q}$ by phonon $\nu$ with a wavevector $\mathbf{q}$. If not specified, mobility and scattering rate are calculated at 300 K. Generally, lower temperature leads to reduced scattering and higher mobility \cite{SI}.

To obtain a converged numerical result from {Eq. (\ref{eqn1})} and {(\ref{eqn2})}, an extremely fine $\mathbf{k}$ and $\mathbf{q}$ sampling over the BZ is required, which makes a direct calculation of the numerous e-ph coupling matrix elements a prohibitive computational burden. Therefore, a Wannier-Fourier interpolation method \cite{2007-PRB-76-165108-Giustino} was used in this study. Electron eigenstates and eigenvalues, vibrational modes and frequencies, as well as e-ph matrix elements were first obtained on a relatively coarse BZ grid ($12\times12\times1$ for $\mathbf{k}$ points and $6\times6\times1$ for $\mathbf{q}$ points). Then they were transformed to the Wannier representation in the corresponding real-space supercells. Electronic Wannier states were determined by the maximally localized Wannier function (MLWF) method \cite{2008-Comput.Phys.Commun-178-685-Mostofi}. Locality in the Wannier representation guarantees that the reverse process from the Wannier representation to the Bloch representation can realize an ultra-dense sampling of the BZ \cite{2016-Comput.Phys.Commun-209-116-Ponce}.

\section{RESULTS AND DISCUSSION}
\subsection{Electron-phonon coupling and intrinsic mobility}

\begin{figure}[tbh]
  \includegraphics[width=8.5cm]{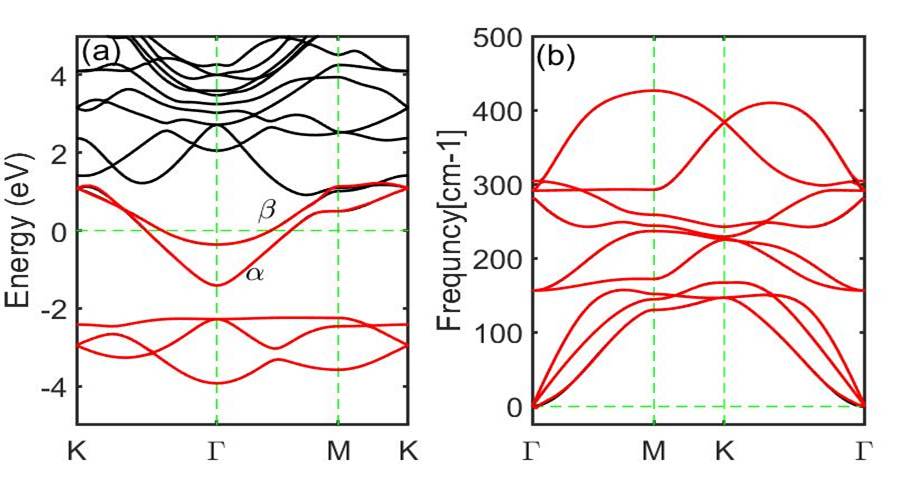}
  \captionsetup{justification=justified}
  \caption[]{\label{band} (a) Electron and (b) phonon band structures of Ca$_2$N monolayer. Results from Wannier interpolation are plotted in red on top of the original results. Bands $\alpha$ and $\beta$ are the 2DEG-FS bands. }
\end{figure}

The electron and phonon band structures of Ca$_2$N monolayer are plotted in {Fig. \ref{band}}. The two bands (labeled as $\alpha$ and $\beta$) crossing the Fermi level which have a quasi-quadratic dispersion around the $\Gamma$ point are the two 2DEG-FS states \cite{2014-JACS-136-13313-Zhao,2017-PRB-95-165430-Inoshita}. Ca$_2$N monolayer has three atoms per unit cell. Therefore, there are nine phonon modes including three acoustic modes and six optical modes. The three acoustic branches are the in-plane longitudinal and transverse acoustic (LA and TA) modes and the out-of-plane acoustic (ZA) mode. The LA and TA branches have a higher frequency than the ZA mode around the $\Gamma$ point. From {Fig. \ref{band}}, we can clearly see a van Hove singularity at the M point.

\begin{figure}[tbh]
  \includegraphics[width=8.5cm]{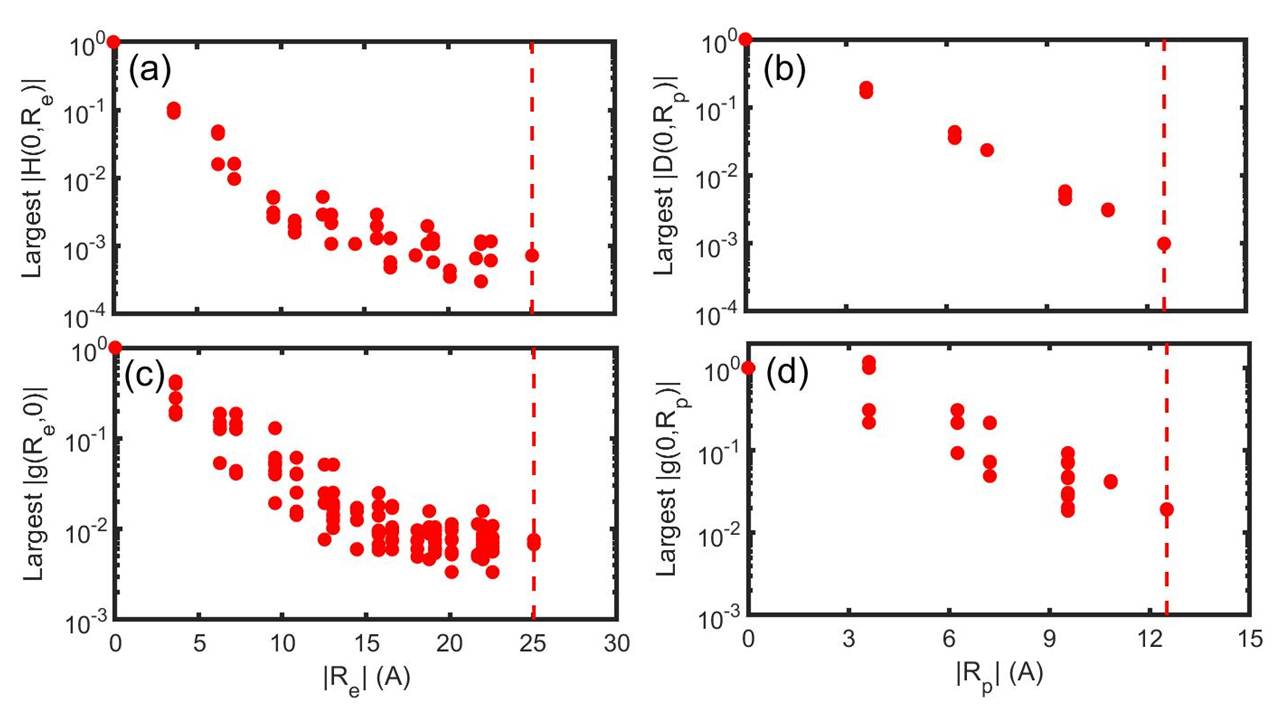}
  \captionsetup{justification=justified}
  \caption[]{\label{decay}Spatial decay of the largest components (only considering absolute values) of (a) electronic Hamiltonian, (b) dynamical matrix, and (c-d) e-ph coupling matrix elements along electron coordinate distance $|\mathbf{R}_e|$ and phonon coordinate distance $|\mathbf{R}_p|$. The data are normalized to their largest values. Truncation distances are marked by dash lines.}
\end{figure}

In {Fig. \ref{band}}, bands calculated via Fourier interpolation from the Wannier representation are plotted in red on top of the bands directly calculated with DFT/DFPT. A good agreement is observed.  Only five electron bands, which are close to the Fermi level, are selected for the Wannier function construction. The success of the Wannier-Fourier interpolation strongly depends on the locality in the Wannier representation \cite{2007-PRB-76-165108-Giustino,2014-JCP-141-034704-Xi,2017-PRB-95-024505-Gao}. As shown in Fig. \ref{decay}, both electronic Hamiltonian and phonon dynamical matrix in the Wannier representation show exponential decay with the distance $|\mathbf{R}|$ between unit cells. Their values at the truncation distance are also small enough compared to those at close distances. In the Wannier representation, e-ph coupling matrix elements are also localized. Therefore, Fourier interpolation can be safely performed to get e-ph coupling matrix elements in the momentum space.

\begin{figure}[tbh]
  \includegraphics[width=8.5cm]{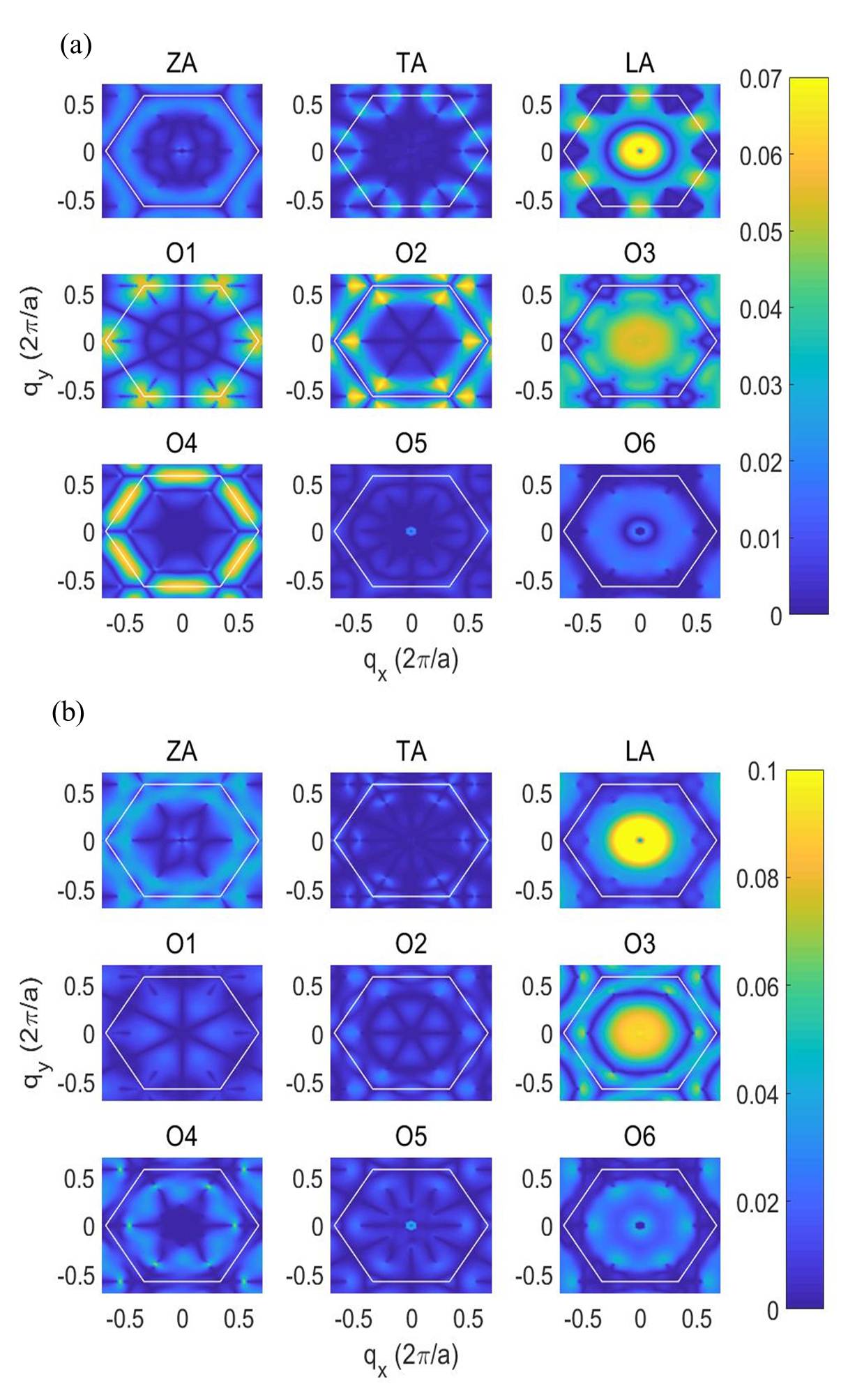}
  \captionsetup{justification=justified}
  \caption[]{\label{ephmat}The e-ph coupling matrix elements $|g_{nm\nu} (\mathbf{k,q})|$ (in unit of eV) as a function of phonon wave vector $\mathbf{q}$, where $\mathbf{k}=\Gamma$ and (a) $n=m=\alpha$ or (b) $n=m=\beta$.}
\end{figure}

The probability of electron transition from one state to another depends on the e-ph coupling matrix elements \cite{2011-PRB-83-161402-Borysenko,2015-WileyInterdiscip.Rev.:Comput.Mol.Sci.-5-215-Xi}. In {Fig. \ref{ephmat}}, we plot the e-ph coupling matrix elements as a function of phonon wavevector for electron scattering from the $\alpha$ or  $\beta$ band at the $\Gamma$ point. A pronounced anisotropy near the BZ boundary is observed. The most strong coupling comes from the LA phonon mode around the zone center. Some optical phonons near the BZ boundary also interact strongly with $\alpha$ or $\beta$ electrons. It is important to notice that e-ph interactions are indeed weak as we expected based on the fact that conducting electrons are form 2DEG-FS.  As we can see from {Fig. \ref{ephmat}}, the largest coupling matrix elements is about 0.1 eV in  Ca$_2$N monolayer, which are notably weaker than those in stanene, silicene, MoS$_2$, silicon, germanium \cite{2017-Adv.Electron.Mater-3-1700143-Nakamura,2013-PRB-87-115418-Li,2015-J.Appl.Phys.-118-045713-Tandon,2015-PRB-92-075405-Li}, noble metals(Cu, Ag, Au) \cite{1979-PRB-19-6130-Varma,2016-PRB-94-155105-Mustafa}, and graphene \cite{2010-PRB-81-121412-Borysenko,2014-NanoLett-14-1113-Park,2016-PRB-93-035414-Gunst}. Notice that this conclusion is not dependent on the choice of the $\Gamma$ state as the initial state for e-ph scattering \cite{SI} and weak e-ph coupling is indeed a general character of 2DEG-FS.

Once the e-ph interaction matrix is calculated, the electron linewidth which is proportional to the scattering rate can be obtained by integrating over the phonon wavevectors in BZ using Eq. (\ref{eqn2}), where the delta functions are treated with a Gaussian smearing. To test the convergence of the electron linewidth with the smearing parameter and the number of $\mathbf{q}$ points, we gradually decrease the smearing parameter $\eta$. For each smearing, enough $\mathbf{q}$ points are used to obtain converged results at this specific smearing.  When the $\mathbf{q}$-point converged results become converged for different smearing $\eta$ values, we obtain a suitable set of $\mathbf{q}$-point number and $\eta$.

\begin{figure}[tbh]
  \includegraphics[width=8.5cm]{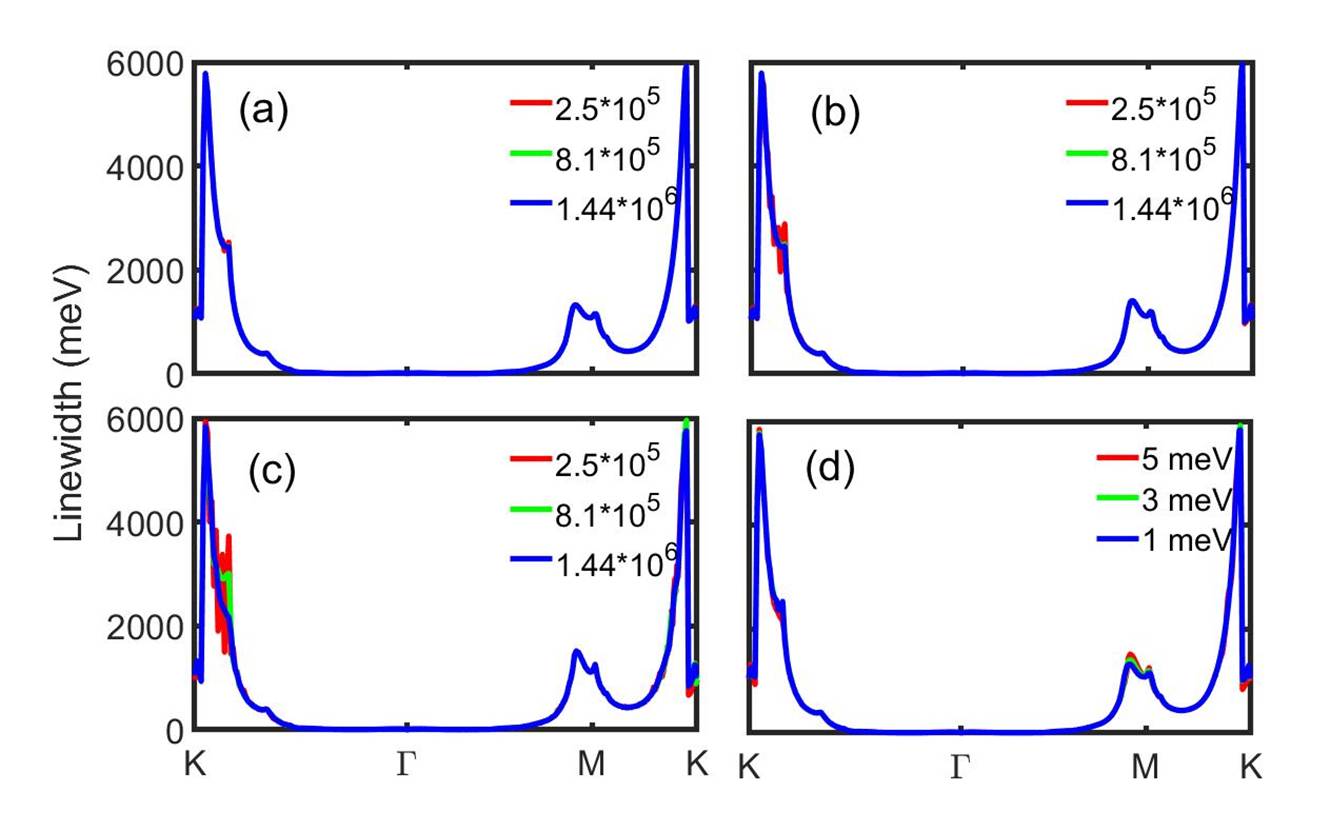}
  \captionsetup{justification=justified}
  \caption[]{\label{linewidth} Electron linewidth for the $\alpha$ band calculated with different smearing (a) 5 , (b) 3, and (c) 1 meV along high symmetry $\mathbf{k}$ points. The numbers of $\mathbf{q}$ points used are indicated with different colors. (d) $\mathbf{q}$-point converged results for different smearing parameters.}
\end{figure}

As shown in {Fig. \ref{linewidth}}, the number of $\mathbf{q}$ points to converge the linewidth for different smearing parameters are different, especially for electronic states near the K point. For a narrow smearing with $\eta$ = 1 meV, we need more than one million $\mathbf{q}$ points to converge the linewidth while a moderate increase of the smearing parameter to 3 meV, for example, can reduce half of the $\mathbf{q}$-point requirement. For the three smearing parameters we tested (5, 3, and 1 meV), the $\mathbf{q}$-point converged linewidth curves are almost the same, indicating that these smearing parameters are already small enough. We employ a 3 meV Gaussian smearing in later calculations, which is expected to give well-converged results.

\begin{figure}[tbh]
  \includegraphics[width=8.5cm]{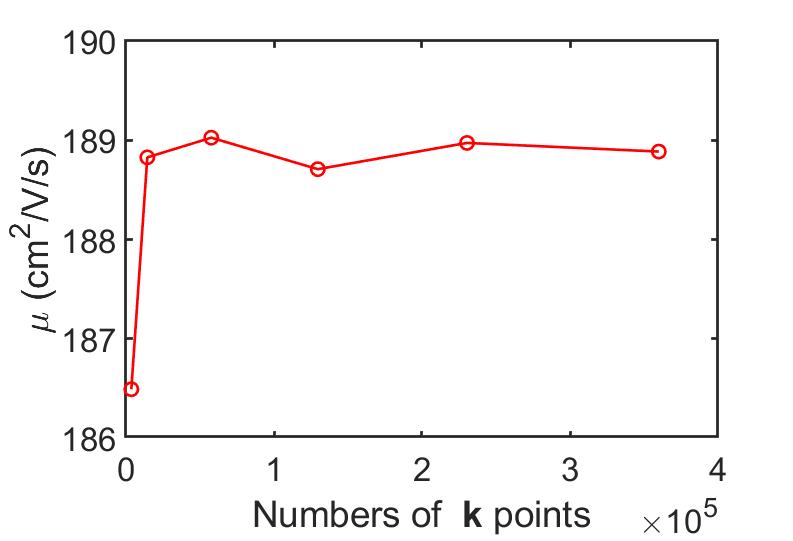}
  \captionsetup{justification=justified}
  \caption[]{\label{mobility} Electron mobility calculated using a 900 $\times$ 900 $\times$ 1 $\mathbf{q}$-point mesh with different numbers of $\mathbf{k}$ points.}
\end{figure}

With a proper smearing parameter and the corresponding number of $\mathbf{q}$-points, we can obtain a well converged relaxation time for each electronic state. Then, the mobility can be calculated by integrating over the BZ using a large number of $\mathbf{k}$ points. As shown in {Fig. \ref{mobility}}, the converged mobility is 189 cm$^2$V$^{-1}$s$^{-1}$. This value is smaller than a rough estimation from bulk Ca$_2$N based on the constant relaxation time \cite{2014-JACS-136-13313-Zhao}. Such a difference indicates that the relaxation time for Ca$_2$N monolayer and bulk Ca$_2$N may be different and it also suggests a possible deficiency of the constant relaxation time approximation \cite{2016-PRB-94-155105-Mustafa}. Notice that the mobility reported here is still significantly higher than that of  common metals \cite{1981-book-8}.

\begin{figure}[tbh]
  \includegraphics[width=8.5cm]{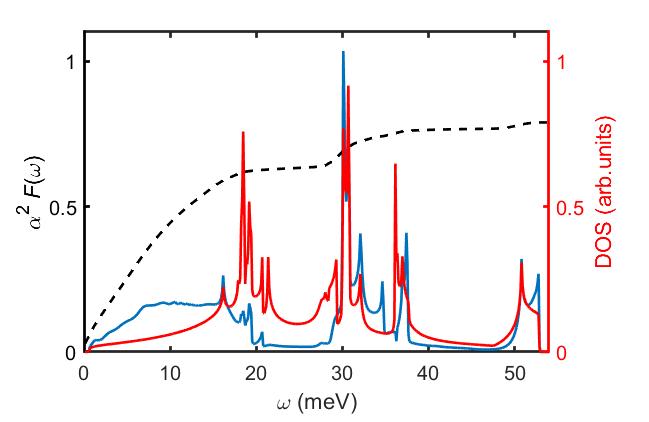}
  \captionsetup{justification=justified}
  \caption[]{\label{superconduct} The Isotropic Eliashberg spectral function $\alpha^2 F(\omega)$ (blue line), phonon DOS (red line), and electron-phonon coupling strength (dash line). The adopted phonon smearing is 0.05 meV.}
\end{figure}

We also calculate the isotropic Eliashberg spectral functions $\alpha^2 F(\omega)$ which is a measure of the e-ph coupling strengths as a function of the phonon frequency \cite{2016-Comput.Phys.Commun-209-116-Ponce}. As shown in {Fig. \ref{superconduct}}, it has a dominant peak around 30 meV. Interestingly, the strength of e-ph coupling $\lambda$ = 0.78, which is much higher than that of graphene (0.21) \cite{2008-NanoLett-8-4229-Park} and even stronger than some ordinary metals (0.42 for Al, 0.16-0.21 for Au, and 0.22 for Na) \cite{1998-PRB-57-11276-Bauer}. Following the McMillan formula \cite{1968-Phys.Rev-167-331-McMillan} within the framework of the BCS theory, we predict that Ca$_2$N monolayer is a BCS superconductor with a transition temperature $T_c$ of 4.7 K under a 0.10 Coulomb pseudopotential \cite{SI}.

\subsection{A comparative study with graphene}

Owing to the 2DEG-FS character of its electronic structure, state-to-state e-ph matrix elements of Ca$_2$N monolayer are small, which leads to a high mobility. However, the intrinsic mobility of Ca$_2$N is still much lower than that of the mostly studied 2D material, graphene (1.67$\times10^5$ cm$^2$V$^{-1}$s$^{-1}$ according to our calculations \cite{SI}), where no 2DEG-FS state exists.  Therefore, it is desirable to perform a comparative study for Ca$_2$N monolayer and graphene on their transport properties.

\begin{figure}[tbh]
  \includegraphics[width=8.5cm]{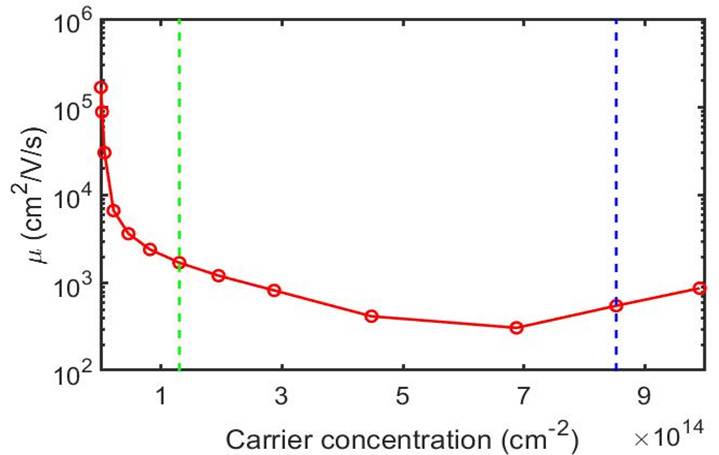}
  \captionsetup{justification=justified}
  \caption[]{\label{doped_graphene} Electron mobility as a function of electron concentration at T = 300 K for graphene. The blue and green dashed lines mark the carrier concentration of neutral Ca$_2$N monolayer and doped Ca$_2$N monolayer with a maximal mobility.  }
\end{figure}

First, we notice that the carrier mobility of graphene strongly depends on its carrier concentration. As shown in  {Fig. \ref{doped_graphene}}, when graphene is doped, its mobility decreases rapidly away from the Dirac point. If we dope graphene to a similar carrier density compared to Ca$_2$N monolayer, we will obtain a mobility larger but in the same order of magnitude compared to that of Ca$_2$N monolayer (555 cm$^2$V$^{-1}$s$^{-1}$ at a carrier concentration of 8.53$\times10^{14}$ cm$^{-2}$). In the following part of this section, when we mention doped graphene we mean graphene doped at this level.

\begin{figure}[tbh]
  \includegraphics[width=8.5cm]{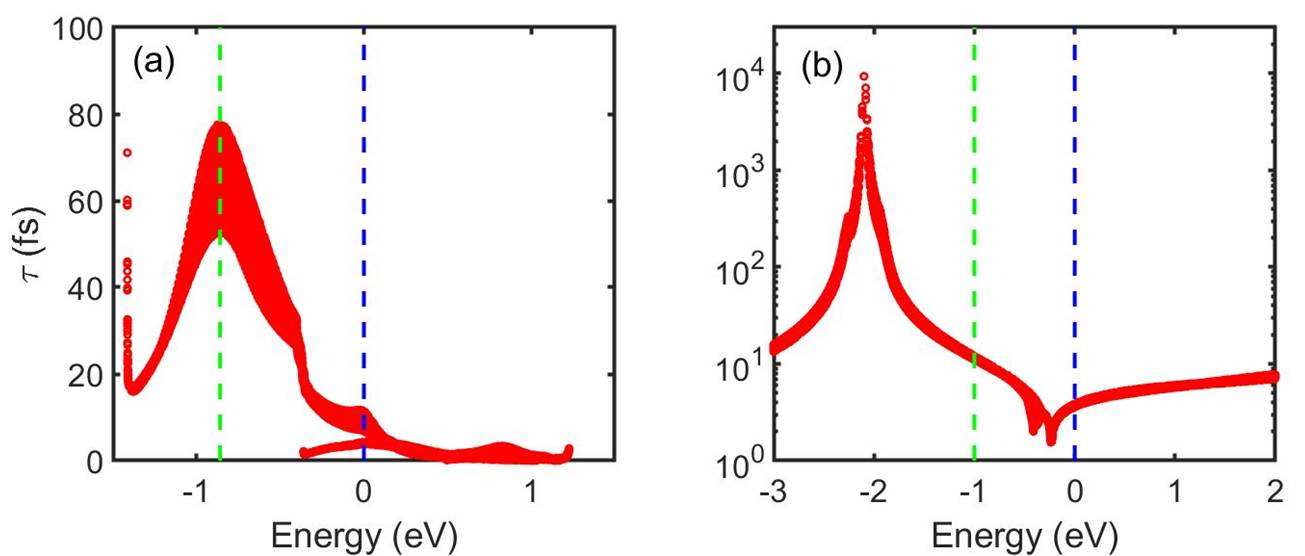}
  \captionsetup{justification=justified}
  \caption[]{\label{relaxation_time} Relaxation times (a) for bands $\alpha$ and $\beta$ in Ca$_2$N monolayer and (b) for doped graphene. The Fermi levels are set to zero.}
\end{figure}

Two important quantities which determine the carrier mobility are relaxation time and group velocity. In {Fig. \ref{relaxation_time}}, we plot the relaxation time as a function of electron energy for both Ca$_2$N monolayer and the doped graphene. For Ca$_2$N monolayer, the two branches come from bands $\alpha$ and $\beta$, respectively. For graphene, the sharp peak is corresponding to the Dirac point, where the relaxation time is beyond 9 ps. At the Fermi level, the relaxation time for electronic states in Ca$_2$N ranges from 3.82 to 11.63 fs while it is from 3.06 to 3.41 fs for doped graphene. Therefore, electrons around the Fermi level have an even bigger chance to be scattered in doped graphene than in Ca$_2$N monolayer.

\begin{figure}[tbh]
  \includegraphics[width=8.5cm]{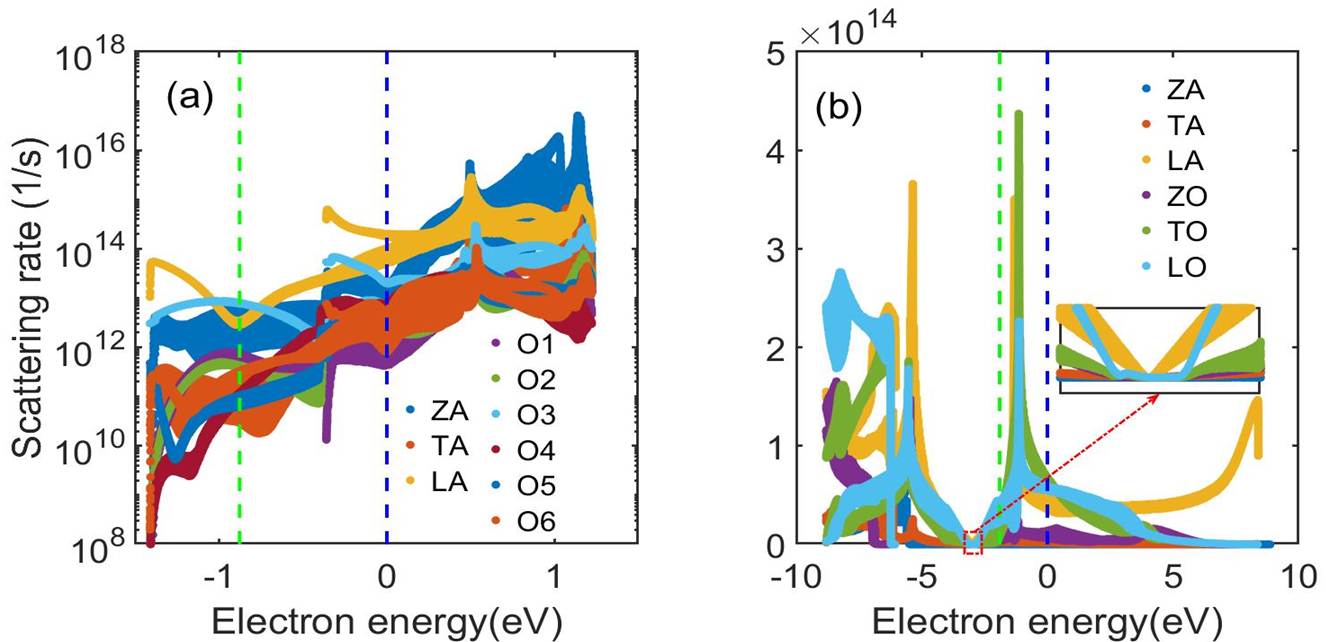}
  \captionsetup{justification=justified}
  \caption[]{\label{scattering_rate} Scattering rate as a function of electron energy for (a) Ca$_2$N monolayer and (b) graphene. The Fermi level is set to zero and different phonon modes are marked.}
\end{figure}

As a further analysis, it is interesting to see which phonon modes make the dominant contribution for e-ph scattering. It turns out that the LA phonon mode is the main scattering source at the Fermi level in Ca$_2$N monolayer ({Fig. \ref{scattering_rate}}), which is consistent with the trend observed in e-ph coupling matrix elements ({Fig. \ref{ephmat}}). Although LA also dominates at the Dirac point, in doped graphene, the LO and TO optical modes dominate the e-ph scattering processes around the Fermi level, a result consistent with previous studies on pristine graphene \cite{2010-PRB-81-121412-Borysenko}. High-frequency optical modes are more difficult to be thermally excited. In fact, the maximum phonon frequency in Ca$_2$N monolayer (429 cm$^{-1}$) is also almost 4 times smaller than that of graphene (1610 cm$^{-1}$  \cite{SI}). Therefore, phonons in graphene are less populated compared to those in Ca$_2$N monolayer, which, considering that the relaxation time in graphene is shorter, further highlights the important role played by the weak e-ph coupling matrix elements due to the 2DEG-FS character in Ca$_2$N monolayer.

\begin{figure}[tbh]
  \includegraphics[width=8.5cm]{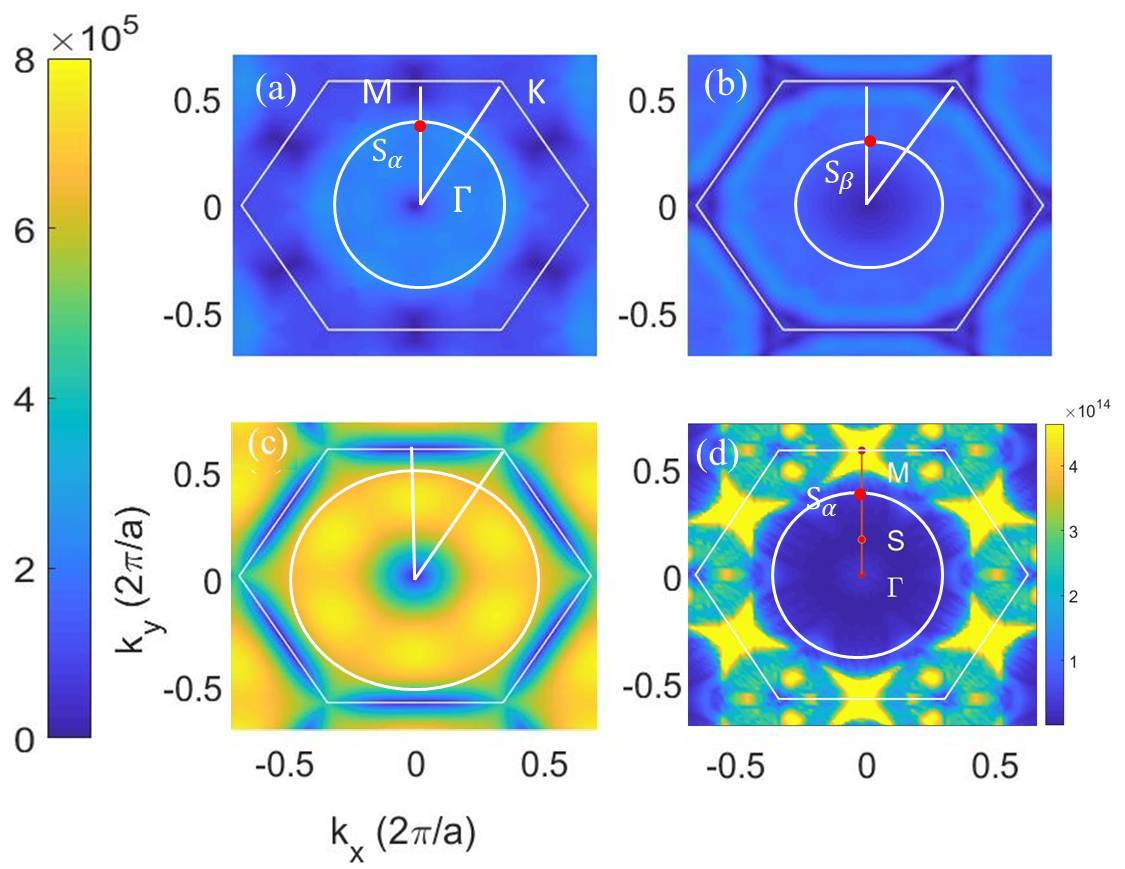}
  \captionsetup{justification=justified}
  \caption[]{\label{velocity} Norm of electron group velocity ($v_{n\mathbf{k}}$) for bands (a) $\alpha$ and (b) $\beta$ of Ca$_2$N monolayer and (c) the conduction Dirac band of graphene. (d) The scattering rate determined by the LA phonon mode for Ca$_2$N monolayer. White circles mark Fermi surfaces. Some special  $\mathbf{k}$-points are marked. }
\end{figure}

Since the 2DEG-FS states have lower scattering rate and longer relaxation time, the electron group velocity is expected to finally determine the carrier mobility which is lower for Ca$_2$N monolayer compared to doped graphene. In {Fig. \ref{velocity}}, we plot the electron group velocity for the bands crossing the Fermi level for both Ca$_2$N monolayer and doped graphene. The Fermi velocity of doped graphene (from 0.39 to 0.43$ \times$10${^6}$ m s$^{-1}$) is higher than that of Ca$_2$N monolayer (0.35 and 0.14$ \times$10${^6}$ m s$^{-1}$ for band $\alpha$ and $\beta$, respectively). The difference becomes even more significant if we consider the energy range within 25 meV from the Fermi energy which is roughly corresponding the thermal fluctuation at 300 K. In the $\pm$25 meV energy range, the group velocity of Ca$_2$N monolayer is within the range from 0.13 to 0.36 $\times$10${^6}$ m s$^{-1}$ while it is 0.38 to 0.59 $\times$10${^6}$ m s$^{-1}$ for doped graphene. Considering that the group velocity appears in Eq. (1) as a quadratic term, such a difference is expected to be enough to explain the mobility difference between  Ca$_2$N monolayer and the doped graphene.

\subsection{Electron concentration dependence of the mobility}
\begin{figure}[tbh]
  \includegraphics[width=8.5cm]{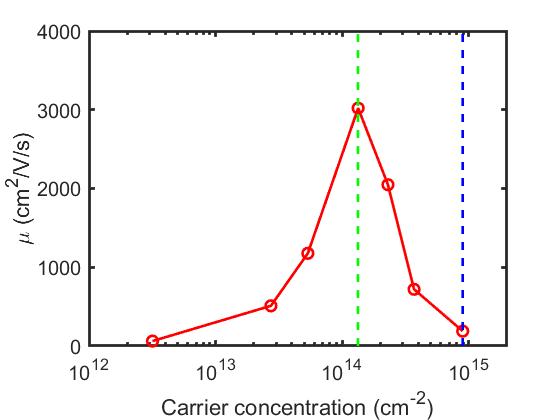}
  \captionsetup{justification=justified}
  \caption[]{\label{concentration} Electron mobility as a function of carrier concentration at T = 300 K for Ca$_2$N monolayer. Carrier concentration is defined as the total number of electrons in bands $\alpha$ and $\beta$. The carrier concentration of intrinsic Ca$_2$N monolayer is 8.93$\times10^{14}$ cm$^{-2}$ as marked by the blue dashed line. The concentration for maximal mobility is 1.33$\times10^{14}$ cm$^{-2}$ which is marked by the green dashed line. }
\end{figure}

It is usually desirable to achieve higher mobility for electronic materials. Different ways to increase mobility have been proposed, including decreasing temperature \cite{2014-JCP-141-034704-Xi} and applying strain \cite{2009-Annu.Rev.Mater.Res.-39-203-Chu}. Here, since we have a strong peak in the relaxation time curve below the Fermi level, it is natural to try to increase the mobility by lowering the electron concentration. As shown in {Fig. \ref{concentration}}, the mobility of  Ca$_2$N monolayer can indeed reach a peak value of about 3000 cm$^2$V$^{-1}$s$^{-1}$ via hole doping, which is an improvement more than one order of magnitude compared to the neutral system and may lead to promising electronic applications. Notice that, although there is a sharp increase of the relaxation time at the band edge of $\alpha$ ({Fig. \ref{relaxation_time}}), there is no mobility maximum there since group velocity there is approaching zero.

According to our phonon mode-resolved scattering rate calculation ({Fig. \ref{scattering_rate}}), one important mechanism for the appearance of the mobility peak is the formation of a valley there in the LA phonon scattering rate. Therefore, in {Fig. \ref{velocity}(d), we plot the scattering rate contributed by the LA phonon mode within the first BZ. The area around the M point where the van Hove singularity locates has the most strong phonon scattering. We pick up three representative $\mathbf{k}$ points, $\Gamma$, S$_\alpha$, and S, for further analysis. S$_\alpha$ is a $\mathbf{k}$ point at the Fermi surface of neutral Ca$_2$N monolayer and the S point corresponds to the mobility peak. As expected, the scattering rate at both S$_\alpha$ (1.32$\times10^{14}$ 1/s) and  $\Gamma$ (9.73$\times10^{13}$ 1/s) is higher than that at S (2.10$\times10^{13}$ 1/s), which is consistent with the result in {Fig. \ref{scattering_rate}(a)}. Interestingly, when we compare the corresponding e-ph coupling matrix elements for initial electronic states located at these three $\mathbf{k}$ points (Fig. \ref{g2}), we find that coupling for S$_\alpha$ states is indeed much stronger but those for $\Gamma$ and S are at the same order of magnitude. Therefore, further analysis is required to explain why electronic states at $\Gamma$ and S have very different scattering rates although the magnitudes of e-ph coupling matrix elements are similar.

\begin{figure}[tbh]
  \includegraphics[width=8.5cm]{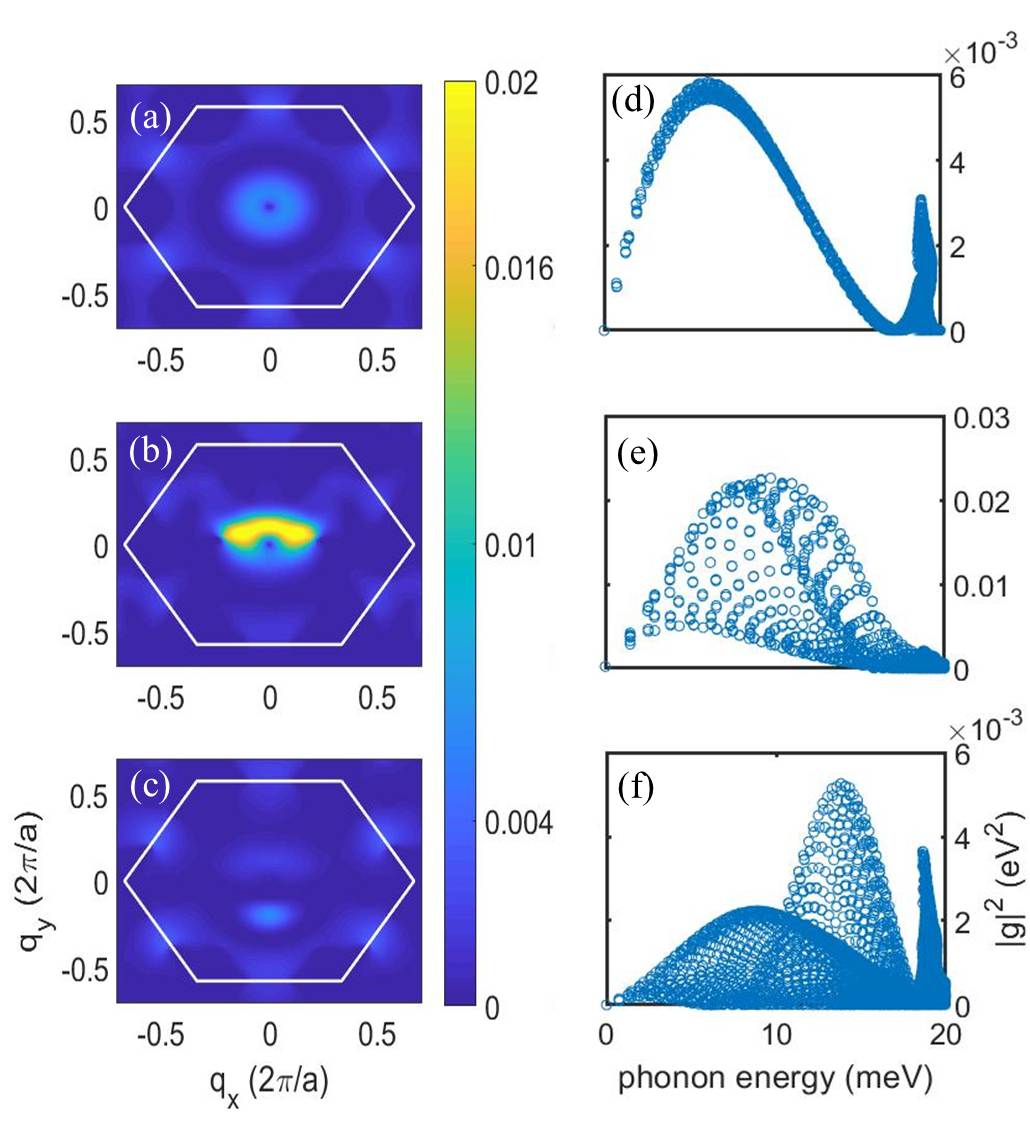}
  \captionsetup{justification=justified}
  \caption[]{\label{g2} Square of e-ph coupling matrix element (in eV$^2$) in Ca$_2$N monolayer as a function of LA-mode phonon wavevectors with the initial electronic states located at the (a) $\Gamma$, (b) S$_\alpha$, and (c) S points. In (d), (e), and (f), these data are projected into different phonon energies.}
\end{figure}

We notice that the phonon momentum space distribution of e-ph coupling is very different for $\Gamma$ and S. In the former case, the e-ph coupling mainly comes from small-wavevector phonons together with some phonons at the M point, which correspond to the two peaks in {Fig. \ref{g2}(d)}. Phonon contribution to the e-ph interaction for S states is much more asymmetric in the BZ, which leads to three peaks in {Fig. \ref{g2}(f)}. Such a difference will lead to different consequences when the energy and momentum conservation law is considered. The $\Gamma$ electron state is mainly scattered by low-momentum phonon, which leads to final electron states also around $\Gamma$. In contrast, the electron state at the S point interacts most strongly with phonon at the -S point in the BZ, which leads to a final state with an $\sim$0.55 eV energy difference with the initial state according to the momentum conservation law. Since the maximal LA phonon energy is only about 0.02 eV, such scattering events are prohibited by the energy conservation law. As a result, the overall scattering rate of the S states is lower than that of the $\Gamma$ states.

With such a phonon momentum mismatch between energy-conserving and strong-coupling states, the mobility of Ca$_2$N monolayer (about 3000 cm$^2$V$^{-1}$s$^{-1}$) at the carrier concentration about  1.33 $\times 10^{14} $ cm$^{-2}$ is already notably higher than that of graphene (about 1700 cm$^2$V$^{-1}$s$^{-1}$). When doped to such a carrier concentration, the Fermi velocities of these two systems are similar. It is 0.34 $\times 10^6$ m/s for Ca$_2$N monolayer and in the range from 0.24 to 0.52 $\times 10^6$ m/s for graphene. In contrast, the scattering rate at the Fermi level is much smaller for Ca$_2$N (about 2 $\times 10^{13}$ 1/s) compared to graphene (about 7 $\times 10^{13}$  1/s). Detailed e-ph coupling matrix element analysis indicates that energy and momentum conservation laws can be easy satisfied in the graphene case because low-momemtum phonons make important contribution in the most relevant optical modes (Fig. S17 \cite{SI}). At the same time, the e-ph coupling matrix elements of graphene are larger than those of Ca$_2$N monolayer. Therefore, it is natural that graphene has a lower mobility at this level of carrier concentration.

\section{CONCLUSION}
In summary, we have investigated the e-ph interaction in the 2D electrene material, Ca$_2$N monolayer, based on first-principles calculations with the Wannier interpolation technique. According to our calculations, Ca$_2$N monolayer has a high mobility of 189 cm$^2$V$^{-1}$s$^{-1}$. Compared to graphene with a similar density of carriers, its scattering rate at the Fermi level is lower, but graphene has higher electron velocities and thus a higher mobility. By reducing its electron concentration, the mobility of Ca$_2$N monolayer can be increased to about 3000 cm$^2$V$^{-1}$s$^{-1}$ via a momentum mismatch mechanism, which outperforms graphene. We have also predicted that Ca$_2$N monolayer becomes a superconductor below 4.7 K. Our results confirm that, as the first electrene system, Ca$_2$N monolayer is a promising electronic material.

\vspace{0.1cm}
\textit{Acknowledgement}: This work is partially supported by NSFC (21222304 and 21573201), by MOST
(2016YFA0200604), by CUSF, and by USTC-SCC, SCCAS, Tianjin and Shanghai Supercomputer Centers.

\bibliographystyle{apsrev4-1}
\bibliography{Ca2N}

\end{document}